# Discrete Logarithmic Fuzzy Vault Scheme

Khaled A. Nagaty

*Abstract*— In this paper a three fuzzy vault schemes which integrated with discrete logarithmic encryption scheme are proposed. In the first scheme, the message $m$ is encoded with discrete logarithmic encryption scheme using randomly generated identity key κ for every message and then divided into non-overlapping segments. In the second scheme, the message is divided into non-overlapping segments and each segment is encoded with discrete logarithmic encryption scheme using the randomly generated identity key κ. In the third scheme, the message is divided into non-overlapping segments where even segments are encoded with identity key $κ_{even}$ and odd segments are encoded with identity key $κ_{odd}$. Identity keys $κ_{even}$ and $κ_{odd}$ are randomly generated for every message. Finally, the encoded segments are declared as coefficients of a polynomial of specific degree. In all proposed schemes, elements of locking set A are used as $X$-coordinate values to compute evaluations of the polynomial by projecting elements of A onto points lying on the polynomial. A large number of random chaff points that do not lie on the polynomial are added to create noise to hide the encoded segments. Security analysis has shown the proposed scheme enjoys provable security over classical fuzzy vaults.

## I. INTRODUCTION

The traditional fuzzy vault scheme is an algorithm operating on a set of elements for locking a message $m$ encoded in the coefficients of a polynomial equation. Its security is based on the difficulty of reconstructing the polynomial equation. Fuzziness is achieved by adding noise points which are related but not identical points called chaff points to the vault which cause confusion to the intruder. The scheme can tolerate some differences between the sets used to lock and unlock the vault but the intruder should face computationally infeasible problems in order to achieve this goal [1-3]. In biometric fuzzy vaults it was shown in [4] that recovering biometric data from templates is possible which motivates us to search for further protection of fuzzy vaults. It is common, in traditional cryptography, if message keys are not identical, the decryption process will output useless random data [5]. So, in order to protect a fuzzy vault from its inherent vulnerabilities traditional cryptographic techniques must be put into action. The existing attacks on fuzzy vaults are categorized into: chaff point attacks where the attack on fuzzy vault schemes depends on the non-randomness observation of chaff points which was observed by Scheirer et al [6]. Chaff points with smaller free area has more neighboring points in the set of fuzzy vault $FV$ than chaff points with larger free area is a shortcoming which makes fuzzy vaults subject to attacks based on chaff points identification [6]. Perturbation attacks are another type of attacks on fuzzy vaults. It is commonly known that perturbation of any bit in a classical cryptosystem key such as AES or RSA will hinder decryption completely. However, perturbation is tolerated in fuzzy vaults which make fuzzy vaults vulnerable against perturbation attacks. Fingerprints fuzzy vaults are vulnerable to correlation attacks which correlate data from two fingerprint impressions of two fuzzy vaults. An intruder uses exhaustive matching to find the best alignment between the two vaults to recover the original elements of the encoding set or the secret polynomial [6]. Correlation is at maximum when the overlap between biometric data of corresponding enrollments is at maximum. Most fingerprints fuzzy vaults that are minutiae-based are vulnerable to brute force attack which breaks the irreversibility requirement. Brute force indicates the security limitation of fuzzy vaults [7]. Even fingerprints fuzzy vaults hardened with passwords are vulnerable to brute force attacks [8-9] such as dictionary attacks where the words in a dictionary are input one after the other. Also, a mix of capital letters, small letters and numbers are used [9]. Hybrid attacks use a dictionary storing hundreds of words with substitutions or appending a character to the beginning or the end of a word [5]. In superstitious key-inversion attack, the attacker obtains a cryptographic secret key of an authorized user by exploiting weak points in the system such as during submitting or transmitting the key [10]. Brute force attack can be improved using statistical attack which exploits the statistical distribution of the feature data. An intruder ranks the vault points using the probability occurrence of corresponding feature data, and samples the points for polynomial interpolation only from the top ranked $w$ points [7]. Using this attack an intruder can insert new points in the available space without violating the minimum distance requirement between points in the vault [11]. Blended substitution attack is the major attack among other types of attacks where the contents of the information in the fuzzy vault can be corrupted with random data whether the intruder has information about the contents or not [12]. For example, in biometric fuzzy vaults during the acquisition of a user's biometric data an attacker can inject a set of biometric data other than the biometric set of the authenticated user. False-accept attack is a serious problem [13-16] in fingerprints fuzzy vaults where the attacker collects a large database contains a pre-aligned minutiae templates that can successively use them to simulate imposter verification. Record multiplicity attacks depend on cross-matching i.e. linking multiple vault records created using the same biometric data. For example, linking two fuzzy vaults created for the same individual using the same genuine points but with different chaff points [17]. Traditional fuzzy vaults contain two drawbacks in the fuzzy commitment scheme which are intolerance of substantial reordering and security over non-uniform distribution.

## II. PREVIOUS WORK

Pioneering work in fuzzy vault schemes was done by Juels and Sudan [1] which was inspired by the previous work of Juels and Wattenberg [3] where error-correcting codes were introduced. Since fingerprints are durable, unchangeable and uniquely identify their owners, this motivated Clancy et al. [18] to use the fuzzy vault scheme to protect the minutiae of fingerprints. Also, U. Uludag, et al. [19] had constructed a fuzzy vault that uses the $x$ and $y$ coordinates of fingerprint minutiae points to secure fingerprint templates. However, traditional fuzzy vaults especially fingerprints fuzzy vaults are vulnerable to several attacks, this motivated K. Nandakumar al. [20] to propose a scheme for hardening a fingerprint fuzzy vault using password but the scheme is also vulnerable against several attacks which motivated Suming Hong et al. [10] to propose a hashed passwords fingerprint fuzzy vault but the problem still exists because passwords are created using: A-Z, a-z, 0-9, and symbols: $, #, +, @, =, /, & hashes of passwords are generated and compared to the given hash. To protect fuzzy vaults against correlation attacks Johannes Merkle et al. [21] proposed a fingerprints fuzzy vault that is correlation resistant and Manvjeet Kaur et al. [22] proposed a fuzzy vault scheme that is combined with Hadamard transformation technique, to prevent correlation attack to an appreciable level. Also, Sungju Lee et al. [23] added chaff points in a more structured way namely 3D lattice to make it computationally hard for an attacker to distinguish between the real points and chaff points of two fuzzy vaults. Blended substitution attack is difficult to prevent, however Sarala S. M. et al. in [12] have launched a blended substitution attack by corrupting the fuzzy vault with random data. By storing the real points of the vault in a server that is connected with the vault's database the fuzzy vault was able to self-recover after the attack. All the above approaches to protect fuzzy vaults against a variety of attacks are based solely on the concept of the difficulty of polynomial reconstruction problem which makes fuzzy vaults vulnerable to all previous attacks. In order for a fuzzy vault to be robust against various attacks conventional cryptographic techniques must take part in the construction of the fuzzy vault. The first attempt to combine cryptographic techniques with the fuzzy vault scheme was proposed by Cun-Zhang Cao et al. [24] where the authors protect a secret key $\kappa$ by dividing it into sub-keys which are encoded into Reed-Solomon codewords before being declared as coefficients of a specific polynomial. T.W. Chim et al. in [25] create a shared key between two parties using the Deffie-Hellman key exchange scheme [26] then they encode this created key into the coefficients of a polynomial of specific degree. The first scheme does not employ the difficulty of solving the discrete logarithmic problem which is believed to be hard with the fuzzy vault scheme which depends on the difficulty of reconstructing a polynomial of specific degree to protect the created shared key. The second scheme uses the Deffie-Hellman scheme to exchange keys but it cannot be used to protect other secret messages. The proposed *discrete logarithmic fuzzy vault* can protect any secret message $m$, where $m$ is firstly encoded with discrete logarithmic encryption scheme then it is hidden in the coefficients of a polynomial of a specific degree. The $discrete\ logarithmic\ fuzzy\ vault$ draws its security from the integration between the difficulties of two problems namely the polynomial reconstruction problem which is believed to be hard in general if $r \ll \sqrt{n.(k-1)}$, where $r$ is the number of real elements, $(k-1)$ is the polynomial degree and $n$ is the vault size [27-31] and solving the discrete logarithm problem. Xavier Boyen in [32] constructed two security models to address the shortcoming of reusing fuzzy messages which makes fuzzy vaults vulnerable to perturbation attack.

The motivation of this work is to enhance the security of the classical fuzzy vault scheme which hides a secret message $m$ into a polynomial of specific degree. The security of the current fuzzy vault depends exclusively on the chaff points that are added to create noise which result in a number of spurious polynomials and does not involve the length of the message $m$ it protects which makes the vault weaker than the message it protects. Therefore, using only locking elements as a method to secure fuzzy vaults are not sufficient without integrating some other techniques into the process. Blending two different security schemes which are based on different cryptographic bases will provide the blended security scheme with the advantages of each security scheme and substitute for the weakness of any scheme with the strength of the other scheme. This is expected to improve the security threshold of protecting a secret message $m$. Hence, the security of the blended *discrete logarithmic fuzzy vault* scheme is based on both a polynomial reconstruction problem and a discrete logarithm problem.

In the proposed blended fuzzy vault scheme, which we call *discrete logarithmic fuzzy vaults* where a randomly generated identity key $\kappa$ for every message is involved in the encryption process. In the first scheme the secret message $m$ is encoded with the identity key $\kappa$ using *discrete logarithmic* encryption scheme. The encoded message is divided into non-overlapped segments and each segment is declared as the coefficient of a polynomial of specific degree. In the second scheme the message $m$ is first divided into $n$ non-overlapped segments where each segment is encoded using *discrete logarithmic* encryption scheme with an identity key $\kappa$. Each encoded segment is then declared as the coefficient of a polynomial of specific degree $k$. In the third scheme odd segments in the $n$ non-overlapped segment are encoded using a randomly generated odd identity key $\kappa_{odd}$ while even segments are encoded using a randomly generated even identity key $\kappa_{even}$. The identity keys $\kappa_{odd}$ and $\kappa_{even}$ are randomly generated for every message. This is in addition to a large number of generated chaff points which create noise to hide the encrypted message $m$ in the vault. The greater the number of such points, the more noise there is to conceal the real polynomial $P(x)$ from an attacker. Blending a classical fuzzy vault which is based on the noise created by a large number of chaff points to hide a secret message $m$ and

*discrete logarithmic* encryption scheme which is based on a cyclic group $H$ of a large prime $p$ and identity keys κ, $κ_{even}$ and $κ_{odd}$ an attacker can get message $m$ if he could solve the discrete logarithm problem on $H$ then reconstructs the polynomial $P(x)$. Hence, the proposed fuzzy vault scheme blends two cryptographic schemes, the first scheme is the *discrete logarithmic* encryption scheme that is directly embedded into the second cryptographic scheme which is the fuzzy vault scheme. For an attacker to access the hidden message $m$ is equivalent to successfully reconstruct the polynomial $P(x)$ and to solve the discrete logarithm problem on $H$. The integrated *discrete logarithmic fuzzy vault* scheme offers an alternative that is provably secure in an information theoretic sense to the well-known fuzzy vault scheme.

This paper is organized as follows: section III is dedicated for traditional fuzzy vault scheme, section IV is dedicated for the proposed *discrete logarithmic fuzzy vault* schemes, section IV is dedicated for non-malleability, section V discusses security analysis and final section VI is dedicated for conclusion.

### III. THE TRADITIONAL FUZZY VAULT SCHEME

The message $m$ to be kept in the vault will be represented in a binary format such that $m \in \{0,1\}^*$ of $l$ bits length. The user that wants to hide the message $m$ in the vault should get his/her finger scanned and a locking set $L$ that contains $x$ and $y$ coordinates of $t$ minutiae points is selected from this finger template. The $x$ and $y$ coordinates of each minutia (8-bits) are concatenated as $X_i = [x_i y_i]$ to obtain a 16-bits locking set $L$. This fingerprint template must have negligible nonlinear distortions and its minutiae are aligned to compensate for the translation and rotation between template and query minutiae data [5]. A finite field $F_q$ is attached to the vault such that:

$$κ' + 1 = \frac{l}{\lceil \log_2(q) \rceil} \quad (1)$$

is the number of elements in $F_q$ which is necessary to encode $m$. It is assumed that $0 < \max_{X \in L} X < q$ and maps $X \to F_q$. Select $f(X) \in F_q[X]$ to be a polynomial of degree $κ \geq κ'$ with coefficients to encode $m$.

The genuine set:

$$G = \{(X_i, Y_i): X_i \in L, Y_i = f(X_i)\} \quad (2)$$

encodes the message $m$. In order to be hard for an attacker to recover message $m$, the genuine points in the genuine set are mixed with chaff points where:

$$C = \{(x_{c_j}, y_{c_j}): X_i \neq X_{c_j}, Y_i \neq Y_{c_j}, i = 1,2,\dots.t,$$
$$j = 1,2,\dots.r - t\} \quad (3)$$

is the set of chaff points where $r$ is the total number of points in the locking set $L$. Therefore, the locking set $L = G \cup C$. Finally, $L$ is passed through a list scrambler which randomizes the list. In order to recover the message $m$ a locking set $U$ containing those $X_i$ values of the vault points [5].

### IV. PROPOSED SCHEMES

In this section, we present our implementations of the three d*iscrete logarithmic fuzzy vault* schemes to overcome the vulnerabilities associated with traditional fuzzy vaults especially biometric fuzzy vaults. Using different polynomials and chaff points for different vaults will not be effective in securing a message $m$ especially if the same lock set $L$ is reused.

#### A. Pre- Lock Stage

This stage is implemented for the lock phase of all the three schemes of *discrete logarithmic fuzzy vault*.

1) Generate a 128-bit signature $S$ using $MD5$ to be appended to message $m$ such that:
$$S = MD5(m) \quad (4)$$
$$m' = [m, S] \quad (5)$$

2) Let $A = \{a_i\}_{i=1}^{t}$ is a locking set containing $t$ elements where $a_i \in F_q$ and $F_q$ is a finite field.

3) Use $α^κ$ where $α$ is a primitive root of a very large prime $p$ which is usually at least 1024-bit to make the fuzzy vault safe and κ is the identity key randomly generated for every message from $\{1 \dots, q-1\}$. For this reason κ is called an ephemeral key.

#### B. Pre- UNLock Stage

This stage is implemented for the unlock phase of all the three schemes of *discrete logarithmic fuzzy vault* where $B = \{b_i\}_{i=1}^{t}$ is an unlocking set where $b_i \in F_q$.

#### C. Algorithm I

*1) Vault Locking*

The concept lies in dividing the message $m'$ into $n$ non-overlapping 256-bit segments $(m'_1, m'_2, \dots, m'_n)$ then each segment $m'_i$ is encoded with *discrete logarithmic* encryption scheme using $α^κ$. Each encoded segment is then declared as a specific coefficient of a polynomial $P$.

- Let $β_i = m'_i α^κ \; \forall \; i = 1,2,\dots,n$ (6)
  Where $m'_i$: is the $i^{th}$ part of the protected message $m'$.
  κ: is the identity key randomly generated for every message $m'$ from $\{1 \dots, q-1\}$.
- Call *Post- Lock*

*2) Vault Unlocking*

To retrieve message $m$ the unlocking set $B$ is used to recover $β_i s$. If $B$ is close enough to $A$, then $B$ will identify most real points in $V$ and $β_i s$ are retrieved.

- If $(|a_i - b_i| \leq δ) \forall i = 1,2,\dots t$ then the corresponding vault point in $V$ is added to the list of points to be used.
- Use Lagrange interpolation to reconstruct polynomial $P$ using $X_v$ and retrieve $β_i^* \forall = 1,2,3,\dots,n$
- If $|β_i - β_i^*| \leq δ$ then $β_i^* = β_i$ (7)
- Multiply $β_i$ by $α^{-κ}$ as follows:
  $β_i α^{-κ} = m'_i α^κ α^{-κ} = m'_i α^0 = m'_i \; \forall i = 1,2,\dots,n$ (8)
- Concatenate $m' \leftarrow [m'|m'_i] \; \forall i = 1,2,\dots,n,$ (9)
  where | means concatenation.
- Call *Post-UNLock*

Algorithm II
*1) Vault Locking*

The concept lies in encoding the message $m$ with *discrete logarithmic* encryption scheme. The encoded message is then divided into $n$ non-overlapping 256-bit segments and each segment is declared as a specific coefficient.

- Let $\beta = m'\alpha^\kappa$          (10)

  Where κ: the identity key is randomly generated for every message $m'$ from $\{1 \ldots, q-1\}$.

  Divide $\beta$ into $(\beta_1, \beta_2, \ldots, \beta_n)$.

- Call *Post-Lock*

*2) Vault Unlocking*

To retrieve message $m$ the unlocking set $B$ is used to recover the $\beta_i s$. If $B$ is close enough to $A$, then $B$ will identify most real points in $V$ and $\beta_i s$ are retrieved.

- If $(|a_i - b_i| \le \delta) \forall i = 1,2,..t$ then the corresponding vault point in $V$ is added to the list of points to be used.
- Use Lagrange interpolation to reconstruct polynomial $P$ using $X_v$ and retrieve $\beta_i^* \forall = 1,2,3,\ldots,n$
- If $|\beta_i - \beta_i^*| \le \delta$ then $\beta_i^* = \beta_i$     (11)
- Concatenate $\beta \leftarrow [\beta|\beta_i] \;\forall i = 1,2,\ldots,n,$     (12)

  where | means concatenation.
- Multiply $\beta$ by $\alpha^{-\kappa}$ as follows:

  $\beta\alpha^{-\kappa} = m'\alpha^\kappa\alpha^{-\kappa} = m'\alpha^0 = m' \;\forall i = 1,2,\ldots,n$   (13)
- Call *Post-UNLock*.

### D. Algorithm III

*1) Vault Locking*

The concept lies in dividing the secret message $m$ into $n$ non-overlapping 256-bit segments then encode the even segments with *discrete logarithmic* encryption scheme using $\alpha^{\kappa_{even}}$ and encode the odd segments using $\alpha^{\kappa_{odd}}$ where $\kappa_{even}$ and $\kappa_{odd}$ are even and odd identity keys randomly generated for every message from $\{1\ldots,q-1\}$ to encode the even segments and odd segments of the secret message $m$ respectively. Each segment is then declared as a coefficient polynomial $P$.

- Let $\beta_i = m_i'\alpha^{\kappa_{even}} \;\forall i = n\%2 = 0, i = 1,2\ldots,n$   (14)

  Where $\kappa_{even}$: the identity key randomly generated for even segments of every message $m'$ from $\{1\ldots,q-1\}$.

- Let $\beta_i = m_i'\alpha^{\kappa_{odd}} \;\forall i = n\%2 \ne 0, i = 1,2\ldots,n$   (15)

  Where $\kappa_{even}$: the identity key randomly generated for odd segments of every message $m'$ from $\{1\ldots,q-1\}$.

- Call *Post-Lock*

*2) Vault Unlocking*

To retrieve message $m$ the unlocking set $B$ is used to recover $\beta_i s$. If $B$ is close enough to $A$, then $B$ will identify most real points in $V$ and $\beta_i s$ are retrieved.

- Consider the unlocking set $B = \{b_i\}_{i=1}^t$ where $b_i \in F_q$.
- If $(|a_i - b_i| \le \delta)\forall i = 1,2,..t$ then the corresponding vault point in $V$ is added to the list of points to be used.
- Use Lagrange interpolation to reconstruct polynomial $P$ using $X_v$ and retrieve $\beta_i^* \forall = 1,2,3,\ldots,n$
- If $|\beta_i - \beta_i^*| \le \delta$ then $\beta_i^* = \beta_i$     (16)
- Multiply $\beta_i$ by $\alpha^{-\kappa_{even}}$ as follows:

  $\beta_i\alpha^{-\kappa_{even}} = m_i\alpha^{\kappa_{even}}\alpha^{-\kappa_{even}} = m_i\alpha^0 = m_i \;\forall i = n\%2 = 0,$
  $i = 1,2\ldots,n$   (17)
- Multiply $\beta_i$ by $\alpha^{-\kappa_{odd}}$ as follows:

  $\beta_i\alpha^{-\kappa_{odd}} = m_i\alpha^{\kappa_{odd}}\alpha^{-\kappa_{odd}} = m_i\alpha^0 = m_i \;\forall i = n\%2 \ne 0,$
  $i = 1,2\ldots,n$
- Concatenate $m' \leftarrow [m'|m_i'] \;\forall i = 1,2,\ldots,n,$   (18)

  where | means concatenation.
- Call *Post-UNLock*.

### E. Post-Lock Stage

This stage is applied to all the three schemes of *discrete logarithmic fuzzy Vault*.

1) Lock $\beta_i s$ in polynomial $P(X)$ with $n-1$ degree such that:

   $P(X) = \beta_n X^n + \beta_{n-1} X^{n-1} + \cdots + \beta_1 X + \beta_0$   (19)
2) $(x_i, y_i) \leftarrow (a_i, P(a_i)) \;\forall i = 1,2,\ldots,t$   (20)
3) $V \leftarrow V \cup (x_i, y_i)$ where $V$ is the set containing real points and chaff points. The set $A$ specifies the $x$-coordinates of real points in $V$, i.e. those points lying on polynomial $P(X)$.
4) Generate chaff points $(u_i, v_i)$ where $u_i \in F - A$ and $v_i \ne P(u_i)$ i.e. not lying on polynomial $P$.
5) $V \leftarrow V \cup (u_i, v_i) \;\forall i = 1,2,\ldots r - t.$   (21)
6) The set $V$ is the union of all real points $(x_i, y_i)$ and all chaff points $(u_i, v_i) \;\forall i = 1,2,\ldots,r$ where $r$ is the total number of real and chaff points.
7) Scramble the fuzzy vault set $V$.

### F. Post-UNLock Stage

This stage is applied to all three schemes of *discrete logarithmic fuzzy Vault*.

1) Extract the message signature $S$ which are the last 128-bits of $m'$ and retrieve the message $m^*$.

2) $S^* = MD5(m^*)$   (22)

3) If $(S^* = S)$ then $m^* = m$   (23)

## IV. NON-MALLEABILITY

To achieve non-malleability it is necessary for the *discrete logarithmic fuzzy vault* user to have a unique $ID$.

*Encoding*:

1) Select random secret key $\kappa$ of *128-bits* length.
2) Choose $64-bits$ $ID$.
3) To identify that the $ID$ of the user is the actual $ID$ generate $16-bits$ CRC data from the user's $ID$.
4) Append the $16-bits$ data to the $64-bits$ $ID$ to construct $IDC$.
5) The $IDC$ is represented as a polynomial of 4 coefficients (64/16) in $GF(2^{16})$ with degree $d=3$ is used to find the coefficients of the polynomial $P$. That is:

   $P(X) = c_3 X^3 + c_2 X^2 + c_1 X + c_0$   (24)
6) Append the 90-*bits* $ID$ to the 128-*bits* secret key $\kappa$ to construct $\kappa ID$ of 208-*bits* of data.
7) The $\kappa ID$ represented as a polynomial of 13 coefficients (192/16) in $GF(2^{16})$ with degree $d=12$ is used to find the coefficients of the polynomial $P$. That is:

   $P(X) = c_{12} X^{12} + c_{11} X^{11} + \cdots + c_1 X + c_0$   (25)
8) Divide $\kappa ID$ into non-overlapping 256-bits segments and each segment is declared as a specific coefficient, i.e. $c_i, i = 0,1,2,\ldots,12$. This mapping from $\kappa ID$ to $c_i$ should be known during decoding as decoded coefficients $c^*_i$ are mapped back to decoded secret $\kappa ID^*$

*Decoding:*
1) Construct the polynomial:
$$P^*(X) = c^*_{12}X^{12} + c^*_{11}X^{11} + \cdots + c^*_1 X + c^*_0 \quad (26)$$
2) Map the coefficients $c^*_i$ to $\kappa ID^*$.
3) The $\kappa ID^*$ is segmented into 2 parts the first 128-bits denote $\kappa^*$ while the remaining 90-bits are the user $ID$ including the $16 - bits$ $CRC$ data.
4) The first $64 - bits$ of the IDC is the reconstructed user $ID^*$ while the remaining $16 - bits$ is the $CRC$ data.
5) To check if this the reconstructed $ID^*$ is the actual $ID$ we divide the polynomial corresponding to $IDC$ with $CRC$ primitive polynomial:
$$P_{CRC}(X) = X^{16} + X^{15} + X^2 + 1 \quad (27)$$
if the remainder is zero with very high probability then we are certain that this is the actual $ID$ of the user and the decoded secret key $\kappa$ is the user's secret, otherwise it is rejected and the reconstructed secret $\kappa^*$ is rejected.

## V. Security Analysis

The security of our proposed *discrete logarithmic fuzzy vault* scheme depends on the difficulty to reconstruct a polynomial $P$ of specific degree $d$ which hides in its coefficients a secret message $m$ after encrypting that message using discrete logarithm encryption scheme. A number of chaff points $r - t$ add noise to increase the difficulty of reconstructing the polynomial $P$. A message signature $S$ is generated using $MD5$ and appended to the end of the message to protect its integrity. One scheme of *discrete logarithmic fuzzy vault* is dividing the secret message $m$ into $n$ segments where each segment $m_i$ is encrypted using a discrete logarithm encryption scheme then each segment is declared a coefficient in polynomial $P$. An alternative scheme is to encrypt the message $m$ using discrete logarithm encryption scheme and then divide the encrypted message into $n$ segments then each segment is declared as a coefficient of polynomial $P$. Message segmentation increases the difficulty for an intruder to retrieve the full message $m$ by decoding and integrating its segments. Dividing the secret message $m$ into $n$ segments makes each segment $m_i$ a discrete logarithm problem which is very hard to be solved for the attacker to reconstruct the full message $m$. The greater the number of message segments the more difficulty for an intruder to retrieve the full message. In addition to this, the security of the proposed fuzzy vault depends on the number of chaff points. The number of chaff points depends on the number of real elements which are directly related to them. The more real elements the more chaff points are added to the fuzzy vault and there will be a set of discrete logarithms that look like $\beta_i$ but are not real. In the absence of additional information the intruder cannot differentiate between the correct $\beta_i$s and the spurious ones. The real $\beta_i$s are hidden with security proportional to the number of spurious discrete logarithms. Let the set $FV$ contains $r$ points such that $r$ are the genuine points and $r - t$ are the chaff points, the probability for an attacker to obtain the correct real points used to lock the genuine polynomial is:

$$\frac{\left(\frac{r}{s}\right)}{\left(\frac{r-t}{s}\right)} = \frac{r}{r-t} \quad (28)$$

which becomes smaller as $r - t$ becomes bigger such that $r < r - t < s$, where $s$ is the number of elements in set $FV$. The probability for an attacker to obtain the genuine polynomial is:

$$\prod_{i=1}^{n} \frac{r}{r-t} \quad (29)$$

which can be written as:

$$\left(\frac{r}{r-t}\right)^n \quad (30)$$

where $n$ is the degree of the genuine polynomial.

## VI. Conclusion

Based on the fuzzy vault scheme and discrete logarithm encryption scheme an integrated discrete logarithm fuzzy vault scheme is proposed. The security of the proposed fuzzy vault is based on both the security of classical fuzzy vault and discrete logarithm security schemes. The applications of the proposed fuzzy vault are the same for the classical fuzzy vault scheme but with higher security level.